\documentclass[conference]{IEEEtran}

\IEEEoverridecommandlockouts

\usepackage{hyperref}
\usepackage{cite}
\usepackage{amsmath,amssymb,amsfonts}
\usepackage{algorithmic}
\usepackage{graphicx}
\usepackage{textcomp}
\usepackage{xcolor}
\usepackage{cleveref}
\usepackage{url}
\usepackage{booktabs}

\renewcommand{\Vec}[1]{\textrm{\boldmath $#1$}} 
\crefname{figure}{Fig.}{Fig.}

\def\BibTeX{{\rm B\kern-.05em{\sc i\kern-.025em b}\kern-.08em
    T\kern-.1667em\lower.7ex\hbox{E}\kern-.125emX}}
\begin{document}
\title{Multi-Sampling-Frequency Naturalness MOS Prediction Using Self-Supervised Learning Model with Sampling-Frequency-Independent Layer\\
}

\author{
\IEEEauthorblockN{Go Nishikawa$^{1*}$, Wataru Nakata$^{1*}$, Yuki Saito$^{1,2}$,\\ Kanami Imamura$^{1,2}$, Hiroshi Saruwatari$^{1}$, and Tomohiko Nakamura$^{2}$}
\IEEEauthorblockA{
$^1$The University of Tokyo, Japan,
$^2$National Institute of Advanced Industrial Science and Technology, Japan.
\\
\{nakata-wataru855, sythonuk\}@g.ecc.u-tokyo.ac.jp ($^*$: equal contributions)
}
}

\maketitle

\begin{abstract}
We introduce our submission to the AudioMOS Challenge (AMC) 2025 Track 3: mean opinion score (MOS) prediction for speech with multiple sampling frequencies (SFs).
Our submitted model integrates an SF-independent (SFI) convolutional layer into a self-supervised learning (SSL) model to achieve SFI speech feature extraction for MOS prediction.
We present some strategies to improve the MOS prediction performance of our model: distilling knowledge from a pretrained non-SFI-SSL model and pretraining with a large-scale MOS dataset.
Our submission to the AMC 2025 Track 3 ranked the first in one evaluation metric and the fourth in the final ranking.
We also report the results of our ablation study to investigate essential factors of our model.
\end{abstract}


\vspace{-5pt}
\section{Introduction}
\vspace{-3pt}
With recent advances in deep neural network (DNN)-based speech synthesis~\cite{ju24naturalspeech3,siuzdak2024vocos,lee2023bigvgan}, fair comparison and evaluation of speech synthesis systems are essential for further development.
The mean opinion score (MOS) test, the gold standard for evaluating the naturalness of synthetic speech, relies on costly and time-consuming annotations by humans.
To mitigate this, DNN-based MOS prediction models~\cite{lo19_interspeech,ldnet,saeki22c_interspeech,baba24utmosv2,cooper22sslmos,le_ssl_mos} have emerged as alternatives to such subjective evaluation.
Widely used models, such as UTMOS~\cite{saeki22c_interspeech}, leverage features from large-scale self-supervised learning (SSL) models~\cite{hsu21hubert,baevski20wav2vec2}, achieving accurate MOS prediction highly correlated with ratings by humans.

One limitation in conventional MOS prediction models is their dependency of sampling frequency (SF) of the backbone SSL models.
Most SSL models are pretrained on speech sampled at a specific SF, such as 16~kHz. Hence, they do not guarantee meaningful feature extraction from speech with different SFs.
Although resampling speech to a specific SF as a preprocessing step can address this issue, it discards high-frequency components, which may contain useful information for predicting the MOS of high-fidelity speech.

In this paper, we present ``MSR-UTMOS'', our MOS prediction model capable of handling speech with multiple SFs, developed for the AudioMOS Challenge (AMC) 2025 Track 3.
This track aimed to accurately predict the results of naturalness MOS test conducted on synthetic speech at 16, 24, and 48~kHz SFs.
Our core contribution is the integration of an SF-independent (SFI) convolutional layer~\cite{saito2022sficonvtasnet} into an SSL model.
We call this architecture ``SFI-SSL model.''
The SFI layer adjusts its weights in accordance with the SF of the input signal, enabling a single MOS prediction model to process speech at various SFs, including those utilizing frequency components higher than the trained Nyquist frequency.
Our submission to the AMC 2025 Track 3 ranked the first in utterance-level mean squared error (MSE) and the fourth in the final ranking based on system-level Spearman's rank correlation coefficient (SRCC).
We also report results of an ablation study to investigate essential factors in our model: SFI-SSL, knowledge distillation (KD) for initializing the parameters of an SFI-SSL model, and fine-tuning the model on MOS prediction.
The code\footnote{\url{https://github.com/sarulab-speech/sfi-utmos}} and pretrained SFI-SSL models\footnote{\url{https://huggingface.co/collections/Wataru/sfi-ssl-6852d2a9382ab3c8b65fc57c}} are available online.

\vspace{-5pt}
\section{AMC 2025 Track 3}\label{sect:amc2025}
\vspace{-3pt}

{\bf General rules:}
This track consists of training and evaluation phases.
During the training phase, participants built their MOS prediction models using the official training set and any publicly available MOS datasets.
The official training set consisted of the results from three separated MOS tests for evaluating synthetic speech at each of the \{16, 24, 48\}~kHz SFs.
The participants could monitor the performance of their models regarding system-level SRCC between the human-annotated and model-predicted MOS on the official validation set.
The official validation set included the same speech samples as the official training set, but MOS for each sample was annotated by a single test that contained speech sampled at \{16, 24, 48\}~kHz.
During the evaluation phase, the organizers first published speech samples from the official evaluation set.
Then, the participants used their models to predict the naturalness MOS on the evaluation set and submitted the results to the leaderboard.
The MOS annotation process for the evaluation set was the same as that for the validation set, but speech samples were unseen during the training phase.

{\bf Evaluation metrics:}
The primal metric, SRCC, measures the rank correlation between the ground-truth ranking and the predicted ranking for compared speech synthesis systems.
In addition, MSE, linear correlation coefficient (LCC), and Kendall's rank correlation coefficient (KTAU) were used as the evaluation metrics.
All metrics were calculated at the utterance and system levels.

{\bf Datasets:}
Each of the training, validation, and evaluation sets contained 400 speech samples.
The 400 samples consisted of 30 utterances from 4 systems for 16~kHz, 30 utterances from 8 systems for 24~kHz, and 5 utterances from 8 systems for 48~kHz.
This setup made the MOS prediction task highly class-imbalanced situation.
Each speech sample was evaluated by 10 listeners, whose IDs were available for building the MOS prediction models.
The speech samples were synthesized by text-to-speech (TTS) and vocoder systems.
The IDs of these systems were shared across the three sets but their details were not disclosed by the organizers.

\vspace{-5pt}
\section{MSR-UTMOS}

\vspace{-3pt}
\subsection{Basic Architecture}
\begin{figure}
    \centering
    \includegraphics[width=0.8\linewidth]{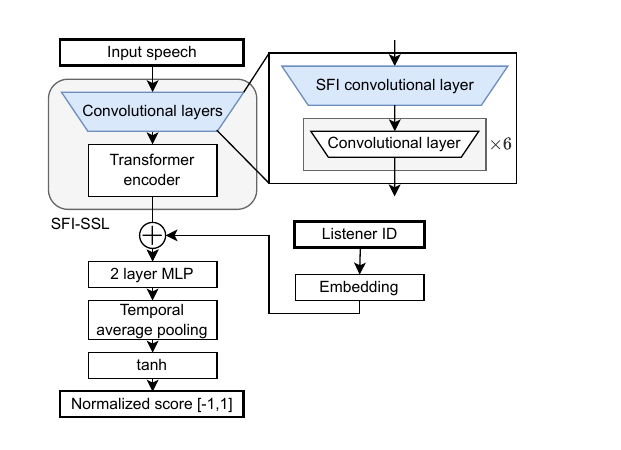}
    \vspace{-4mm}
    \caption{Architecture of the proposed MOS prediction system.}
    \label{fig:model}
\end{figure}

\vspace{-3pt}
Figure~\ref{fig:model} shows the basic architecture of MSR-UTMOS.
It is built upon SSL-MOS~\cite{cooper22sslmos} that uses a pretrained SSL model as a feature extractor from input speech.
In addition to the listener-ID conditioning~\cite{saeki22c_interspeech,ldnet,le_ssl_mos}, we investigate the effectiveness of the SFI convolutional layer~\cite{saito2022sficonvtasnet} in the MOS prediction with multiple SFs.
Concretely, we replace the ordinary convolutional layer in the SSL model with its SFI counterpart and name this backbone model ``SFI-SSL.''

\vspace{-3pt}
\subsection{SFI Convolutional Layer}
\begin{figure}[t]
    \centering
    \includegraphics[width=0.8\linewidth]{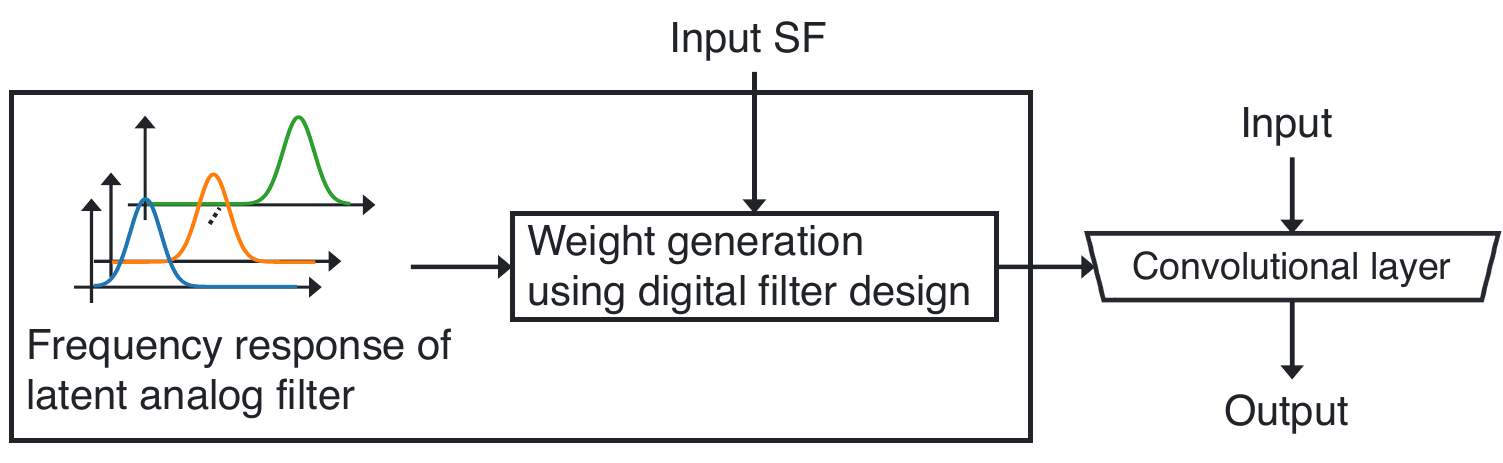}
    \vspace{-4mm}
    \caption{Schematic of SFI convolutional layer.}
    \label{fig:sfi_conv}
\end{figure}

\vspace{-3pt}
The SFI convolutional layer is an extension of the convolutional layer for handling arbitrary SFs. 
This layer is based on the fact that the weights of the convolutional layer can be interpreted as the collection of digital filters.
From this interpretation, this layer generates the weights from a function representing the frequency response, which we call the latent analog filter.
By utilizing digital filter design method, the generated weights are consistent for any SFs.
Thus, this layer can handle arbitrary SFs.
See~\cite{saito2022sficonvtasnet} for details of the digital filter design method.
\Cref{fig:sfi_conv} shows the schematic of the SFI convolutional layer.
For the latent analog filter, we use neural analog filter (NAF), which represents the frequency response via a DNN~\cite{imamura2024naf}.
The NAF consists of an input transformation method based on random Fourier features (RFFs)~\cite{tancik2020rff} and a simple feedforward network.

The SFI convolutional layer was originally introduced to audio source separation~\cite{saito2022sficonvtasnet}.
Concretely, the SFI layers were used at the beginning and end of DNNs for separation.
Because the goal of the AMC 2025 Track 3 is the MOS prediction that can handle speech with various SFs, we simply introduce the SFI layer at the beginning of our MOS prediction model, as shown in Fig.~\ref{fig:model}.
To our best knowledge, this is the first study to introduce the SFI convolutional layer to the SSL model and investigate its effect on the MOS prediction.

\vspace{-3pt}
\subsection{KD from Non-SFI-SSL Model for SFI-SSL Model}
\vspace{-3pt}
Training SSL models from scratch can be computationally expensive. To mitigate this, we incorporate KD~\cite{Hinton2015DistillingTK} from a pretrained non-SFI-SSL model into the initialization for our SFI-SSL model. 

Let $\Vec{x}^{(f_s)}$ be input speech sampled at $f_s$.
We first resample $\Vec{x}^{(f_s)}$ at two different SFs: 1) 16~kHz to obtain $\Vec{x}^{\rm (16k)}$ for input to a non-SFI-SSL model and 2) randomly selected SF $f_u$ from \{8, 16, 24, 48\}~kHz for the SFI-SSL model under the condition $f_s \geq f_u$. 
For each resampled speech, we extract the feature representations from the non-SFI-SSL and SFI-SSL models, i.e., $\Vec{h} = f(\Vec{x}^{\text{(16k)}}; \theta)$ and $\Vec{h}_{\text{SFI}} = g(\Vec{x}^{(f_u)};\theta_{\text{SFI}})$, where $\theta$ and $\theta_{\text{SFI}}$ represent the model parameters.
For example, $\Vec{h}$ from an $L$-layer SSL model consists of $[\Vec{h}^{(1)}, \ldots, \Vec{h}^{(L)}]^\top$, where $\Vec{h}^{(l)}$ is extracted from the $l$-th layer of the model.
The KD loss $\mathcal{L}_{\text{KD}}$ is formulated as the feature matching between the non-SFI-SSL (i.e., fixed teacher) and SFI-SSL (i.e., trainable student) models and defined as:
\begin{align}
    \mathcal{L}_{\text{KD}} &= \sum_{l=1}^{L} \| \Vec{h}^{(l)} - \Vec{h}_{\text{SFI}}^{(l)} \|_2^2.
\end{align}
By minimizing this loss, we aim to distill rich semantic knowledge learned by pretrained non-SFI-SSL models to SFI-SSL models.

\vspace{-3pt}
\subsection{MOS Prediction Network and Its Training Strategy}
\vspace{-3pt}
The SFI-SSL model initialized with the KD loss minimization is then fine-tuned for MOS prediction.
To improve the final performance, we also introduce the listener-ID conditioning~\cite{saeki22c_interspeech}.
Concretely, the output from the last layer of the SFI-SSL model $\Vec{h}_{\text{SFI}}^{(L)}$ is summed to the listener embedding and then fed into the 2-layer neural network to predict the listener-specific score.
The whole model is trained to minimize the MSE between the listener-conditioned ground-truth score and the predicted score.
Following the previous work~\cite{saeki22c_interspeech}, we make the target score $\tilde{S}$ normalized within the range $[-1, +1]$ by
$\tilde{S} = (S - 2) / 3$ where $S \in [1, 5]$ is the unnormalized score.
Following previous work~\cite{baba24utmosv2}, we also pretrain the MOS prediction model on a larger MOS dataset than the official training dataset to enhance the generalizability of the model.

\section{Experiments}
\vspace{-3pt}
\subsection{Baseline model}
\vspace{-3pt}
The baseline model was identical to those of the proposed MOS prediction model except for using the non-SFI-SSL models and SF conditioning.
For the given SF from \{16, 24, 48\}~kHz, we prepared an embedding layer to encode this SF information.
The model predicted MOS from the sum of the SF embedding, listener embedding, and features from a non-SFI-SSL model to perform SF-adaptive, listener-conditioned MOS prediction.

\vspace{-3pt}
\subsection{Experimental conditions}
\vspace{-3pt}
\textbf{Datasets:}
For KD-based initialization of the proposed SFI-SSL models, we used all speech samples from EARS~\cite{richter2024ears}, a 100 hours of 48~kHz sampled English speech by 107 speakers.
For fine-tuning the non-SFI- and SFI-SSL models on the MOS prediction, we used the train subset of BVCC~\cite{cooper21_ssw} containing the naturalness scores of 4,974 synthesized speech.
Each of the samples is associated with naturalness ratings from 8 listeners, thus the total number of ratings is 39,794 ratings.
For the final training, we used the AMC 2025 Track 3 official sets.

\textbf{SSL models:}
We adopted three widely-used SSL models: wav2vec2.0 (w2v2)~\cite{baevski20wav2vec2}, HuBERT~\cite{hsu21hubert}, and WavLM~\cite{chen22wavlm} as the backbone feature extractor and prepared their non-SFI and SFI versions.
As the initialized models, we used \href{https://huggingface.co/facebook/wav2vec2-base}{\texttt{wav2vec2-base}}, \href{https://huggingface.co/facebook/hubert-base-ls960}{\texttt{hubert-base-ls960}}, and \href{https://huggingface.co/microsoft/wavlm-base}{\texttt{wavlm-base}} available on HuggingFace.
All of them were pretrained on LibriSpeech~\cite{panayotov2015librispeech} sampled at 16~kHz.

\textbf{DNN architectures:}
For SFI convolutional layer, we used a frequency domain NAF with maximum frequency of 24~kHz.
We used the same network architecture for the NAF as in~\cite{imamura2024naf}.
To maintain the 50 Hz frame rate of the non-SFI-SSL models, for each of the \{16, 24, 48\}~kHz SFs, we set the (kernel size, stride) to (10, 5), (15, 7.5), and (30, 15), respectively.
To handle the non-integer stride, we used the interpolation-based algorithm for the convolutional layers~\cite{imamura2023noninteger}.
For the MOS prediction network, we used multi-layer perceptrons with 768 feature dimension with SiLU~\cite{Elfwing2017SigmoidWeightedLU} activation.  

\textbf{DNN optimization:}
We used schedule-free version of RAdam optimizer~\cite{defazio2024the,Liu2020On} with its hyperparameters setting to $\eta=1\times 10^{-4}$, $\beta_1 = 0.9$, and $\beta_2 = 0.999$. 
We set the batch size to 32 for the KD, 64 for the pretraining, and 24 for the final training.
The KD was performed for one day with single H200 GPU. 
The pretraining and final training were performed for 50 epochs, which was equivalent to 8 hours with 1/7-th of H200 GPU. 
We selected the best checkpoints based on the MSE on the validation set of BVCC for pretraining. 
For the final training, the checkpoint with minimum loss on the validation set acquired by cross validation were selected.
Note that this validation set was different from the Track 3 official validation set.

\textbf{Evaluation:}
We calculated MSE and SRCC on the utterance level and system level, using the Track 3 official evaluation set.
Upon inference, we first performed listener-wise score prediction using the IDs of 10 listeners in the Track 3 official training set.
Then, we averaged the predicted scores as MOS.
We repeat this process for seven models from each cross-validation fold and the final prediction was the average of those predicted MOS.

\vspace{-3pt}
\subsection{Results}
\vspace{-3pt}
\textbf{Challenge Results:}
We describe the performance of our model in the AMC 2025 Track 3, where seven teams submitted their scores.
For the challenge submission, we took the average of the predictions from our models with the SFI versions of w2v2, HuBERT, and WavLM.
The result of our model is presented in \cref{tab:my_label} as ``T19.''
Our system ranked the first in utterance-level MSE and the fourth in system-level SRCC.
\begin{table}[t]
    \centering
    
    \caption{Evaluation results. For each column, worst values are indicated with \underline{underline} and the best values are indicated with \textbf{bold} (``SFI'': with SFI-SSL, ``PT'': with pretraining on BVCC, ``T19'': submission to the challenge).}
    \label{tab:my_label}
    \begin{tabular}{l|cc|rr|rr}
    \toprule
     & & & \multicolumn{2}{c}{System-level} & \multicolumn{2}{c}{Utterance-level} \\
    Model & SFI & PT & MSE & SRCC & MSE & SRCC \\
    \midrule
    w2v2     & \checkmark & \checkmark   & 0.081 & \textbf{0.920} & 0.234 & 0.702 \\
             &            & \checkmark   & 0.092    & 0.851 & \textbf{0.214} & \textbf{0.741} \\
             & \checkmark &              & 0.091 & 0.851 & \underline{0.300} & 0.648 \\
                             &            &              & 0.095    & 0.817 & 0.258 & 0.686 \\\midrule
    HuBERT & \checkmark & \checkmark   & 0.085 & 0.902 & 0.259 & 0.673 \\
     &            & \checkmark   & 0.095 & 0.913 & 0.247 & 0.695 \\
     & \checkmark &              & 0.104 & 0.857 & 0.276 & 0.660 \\
                       &            &              & 0.103 & \underline{0.803} & 0.275 & 0.672 \\\midrule
    WavLM   & \checkmark & \checkmark   & 0.078 & 0.907 & 0.250 & 0.684 \\
      &            & \checkmark   & \textbf{0.073} & 0.830 & 0.243 & 0.680 \\
      & \checkmark &              & 0.082 & 0.884 & 0.270 & \underline{0.644} \\
                       &            &              & \underline{0.119} & 0.841 & 0.279 & 0.673 \\\midrule
    T19 & \checkmark & \checkmark   & 0.080 & 0.914 & 0.238 & 0.694 \\
    \bottomrule
    \end{tabular}
\end{table}

\textbf{Ablation study:}
\Cref{tab:my_label} shows the results of our ablation study.
We can see that ``w2v2 w/ SFI w/ pretrain'' achieves the best system-level SRCC of 0.920.
This result demonstrates the effectiveness of our SFI-SSL model for MOS prediction with multiple SFs.
We can also find that the pretraining on the large MOS dataset tends to improve the evaluation metrics, which is inline with the previous studies~\cite{baba24utmosv2}

However, for utterance-level metrics, the best performing model was ``w2v2 w/ pretrain.''
This result suggests that, although the SFI-SSL could improve the {\it system-level} MOS prediction performance, it has room for improvement for {\it utterance-level} MOS prediction.
We plan to introduce more robust MOS prediction framework such as the contrastive loss minimization~\cite{saeki22c_interspeech} considering the SF difference into our model.

\vspace{-5pt}
\section{Conclusion}
\vspace{-3pt}

This paper introduced our MOS prediction model  MSR-UTMOS which we submitted to the AMC 2025 Track 3.
The results confirmed the effectiveness of the SFI convolutional layer for MOS prediction with multiple SFs.
In future work, we will explore the applicability of SFI-SSL models to other speech processing tasks.

\textbf{Acknowledgements:}
This work was supported by JST Moonshot JPMJMS2011, JST BOOST JPMJBY24C9, JSPS KAKENHI JP23K28108 and Research Grant S of the Tateisi Science and Technology Foundation.

\bibliographystyle{bib/IEEEbib}
\bibliography{bib/refs}

\end{document}